\documentclass[review]{elsarticle}
\usepackage{lipsum}
\makeatletter
\def\ps@pprintTitle{%
 \let\@oddhead\@empty
 \let\@evenhead\@empty
 \def\@oddfoot{}%
 \let\@evenfoot\@oddfoot}
\makeatother
\usepackage{lineno,hyperref}
\modulolinenumbers[5]
\usepackage[utf8]{inputenc}
\usepackage[T1]{fontenc}
\usepackage{mathtools}
\usepackage{subfig}
\usepackage{float}
\usepackage{graphicx}
\usepackage[autostyle]{csquotes}
\usepackage[pdftex]{color}
\begin{document}
\begin{frontmatter}
\title{Reconstructing complex field through opaque scattering layer with structured light illumination}


\author[mymainaddress,mysecondaryaddress]{Aditya Chandra Mandal}
\author[mymainaddress]{Manisha}
\author[mythirdaddress]{Abhijeet Phatak}
\author[myfourthaddress]{Zeev Zalevsky}
\author[mymainaddress]{Rakesh Kumar Singh\corref{mycorrespondingauthor}}

\cortext[mycorrespondingauthor]{Corresponding author}
\ead{krakeshsingh.phy@iitbhu.ac.in}

\address[mymainaddress]{Laboratory of Information Photonics and Metrology, Department of Physics, Indian Institute of Technology (Banaras Hindu University), Varanasi, 221005, Uttar Pradesh, India}
\address[mysecondaryaddress]{Department of Mining Engineering, Indian Institute of Technology (Banaras Hindu University), Varanasi, 221005, Uttar Pradesh, India}
\address[mythirdaddress]{330 N Mathilda Ave 201 Sunnyvale CA, USA - 94085}
\address[myfourthaddress]{Bar-Ilan University, Faculty of Engineering and Nano Technology Center, Ramat-Gan, Israel}

\begin{abstract}
The wavefront is scrambled when coherent light propagates through a random scattering medium and which makes direct use of the conventional optical methods ineffective. In this paper, we propose and demonstrate a structured light illumination for imaging through an opaque scattering layer. Proposed technique is reference free and capable to recover the complex field from intensities of the speckle patterns. This is realized by making use of the phase-shifting in the structured light illumination and applying spatial averaging of the speckle pattern in the intensity correlation measurement. An experimental design is presented and simulated results based on the experimental design are shown to demonstrate imaging of different complex-valued objects through scattering layer.
\end{abstract}


\end{frontmatter}

\section{Introduction}
Extracting the complex wavefront of the light is desired in various fields, including diffraction, microscopy, bio-medical optics, and imaging through random scattering medium \cite{goodman2015statistical,j1984laser,somkuwar2017holographic,liu2022robust,escobet2018wide}. Phase information of the light is hard to be directly measured due to the extremely high frequency of the optical field, and only the amplitude information is directly observed from the available sensors. In contrast, variety of coherent light applications demands simultaneous sensing of the amplitude and the phase of light field. Availability of the complex field information, i.e., amplitude and phase, supports the coherent propagation of the light and permits digital recovery and reconstruction of the information at any arbitrary plane under paraxial condition. Importance of coherent field propagation has been highlighted in the microscopy, imaging through random scattering medium, and analysis of the optical fields. The complex field measurement is broadly categorized into two groups.  The first category derives its basic principle from the geometrical optics and a common technique in this category is the Shack-Hartman wavefront sensor which uses micro lens arrays\cite{gong2016holographic,cha2010shack, tyson2022principles}. A new development  in the wavefront sensing is reported by replacing the lens arrays with the spatial light modulator and applying the computational method to recover the phase\cite{wang2017wavefront,mico2009phase}. Phase recovery by measurement of multiple intensity measurements and iterations have also been reported\cite{yoneda2021single,zuo2020transport,gupta2020transport,maallo2010quantization,almoro2009angular}. The second category covers interferometric methods to extract the complex field of the light\cite{park2018quantitative,mico2019resolution,lin2022deep}. The interferometric methods offer high-resolution phase recovery without iterations and are widely utilized for various quantitative phase imaging and non-destructive applications. However, interferometry demands  a reference and hence makes the experimental system bulky and vulnerable to the external disturbances. Moreover, these wavefront sensing methods are mainly tested for the homogenous media.\\
\\
On the other hand, in many applications, it is required to extract the information from a randomly scattered media. The randomly scattering may arise as a result of light propagation through an inhomogeneous refractive medium such as scattering wall, multimode fiber, turbulence, etc. Propagation of a coherent light through a random scattering medium generates a complicated pattern known as laser speckle\cite{goodman2015statistical}. When the object is situated behind the random scattering media, the conventional optical techniques are not suitable to extract the useful information directly from the speckle pattern. Different techniques have been developed to reconstruct the desired information from the speckle. A set of techniques to suppress the randomness applies  adaptive optics\cite{tyson2022principles}, phase conjugation\cite{lee2015one,hillman2013digital}, wavefront shaping\cite{sanjeev2019non,sanjeev2022optical}, optimization methods\cite{katz2014non,katz2012looking}, correlation optics\cite{rosen2022roadmap,kumar2014recovery,PhysRevApplied.13.044042,PhysRevLett.113.263903,chen2021wavefront}, etc. Other methods utilize a linear and a deterministic relation between the incident and the output wavefield to deliver information through a random scattering medium\cite{kim2015transmission,van2010information,yoon2015measuring}. Projection of an unknown speckle pattern on the object can be used to enhance the spatial resolution of a diffraction limited system\cite{wagner2015superresolved,vinu2019speckle}.\\
\\
A scattered optical fields preserve a certain degree of correlation and the correlation-based techniques have recently enabled several new methods for imaging through random scattering medium. Speckle correlation related to the memory effect brings an advantage in the non-invasive imaging of the target through the random scattering medium\cite{katz2014non,bertolotti2012non}. These methods utilize correlation with the phase retrieval algorithms to see through a scattering medium. In contrast, the Hanbury Brown-Twiss (HBT) approach along with the holographic approach is utilized in recent years to resolve the phase loss problem in the speckle correlation\cite{kumar2014recovery}. In a series of communications, we have reported different techniques for imaging through diffuser using the HBT approach\cite{chen2022increasing,singh2017hybrid,chen2020phase,kim2019imaging,singh2014quantitative}. Basic principles of these techniques originate from interference of the coherence waves in the intensity correlation and these techniques are free from iterations and convergence issues of the phase retrieval algorithm. 
However, the interferometry makes the system bulky and demands a desired reference coherence function to get the interference fringes in the cross-covariance function\cite{mandal2022structured}.\\
\\
In this paper, we introduce a structured light illumination to develop a new non-interferometric technique for digitally reconstructing the complex field through opaque scattering layer. Structured light illumination has received significant attention in recent years and majority of these interests consider propagation through a free space or homogeneous media, except for some limited studies \cite{fu2021light,lai2018coded,kashter2016enhanced,wen2021structured,hou2021complex,zhang2015single,martinez2017single,chen2021time}. Here, we propose and first time apply, to the best of our knowledge, structured light illumination for digital recovery of complex field through opaque scattering layer. This is achieved by illuminating the target obstructed by opaque scattering layer with structured light and applying a intensity correlation measurement. We also propose a new scheme of recovering information from the intensity correlation by spatial ergodicity (rather than rotating diffusers). This strategy helps to apply proposed method for looking through a static diffuser and overcoming phase loss problem in the conventional HBT approach. As an application of our technique, a complex Fourier spectrum from the intensities of the random speckle patterns is successfully obtained and a complex-valued object is reconstructed. A detailed theoretical framework of the proposed method, a possible experimental scheme to implement this work, and simulation results based on the experimental model are presented in the coming section.

\section{Theory}
A schematic of the existing challenge and its solution by the structured light illumination is highlighted in Figs. 1(a) and 1(b) respectively.The object information, scrambled by the scattering layer as shown in Fig. 1(a), is Fourier transformed by a lens and detected at the observation plane.

\begin{figure}[!ht]
  \centering
  \begin{tabular}[b]{c}
    \includegraphics[width=8cm]{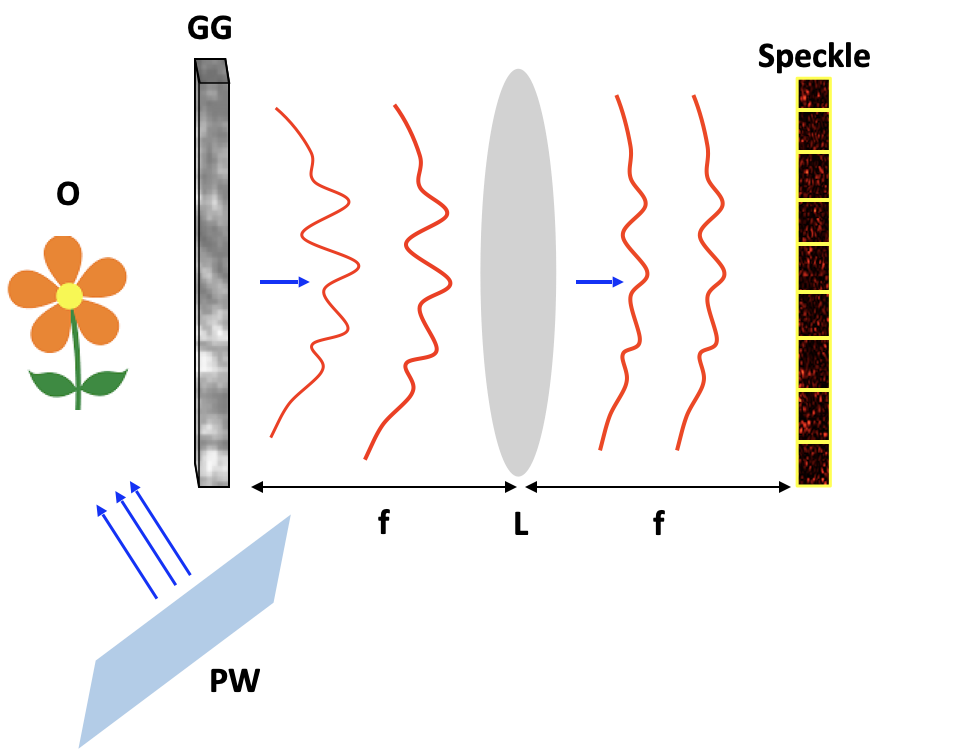} \\
    \small (a)
  \end{tabular} \enspace\\
  \begin{tabular}[b]{c}
    \includegraphics[width=10cm]{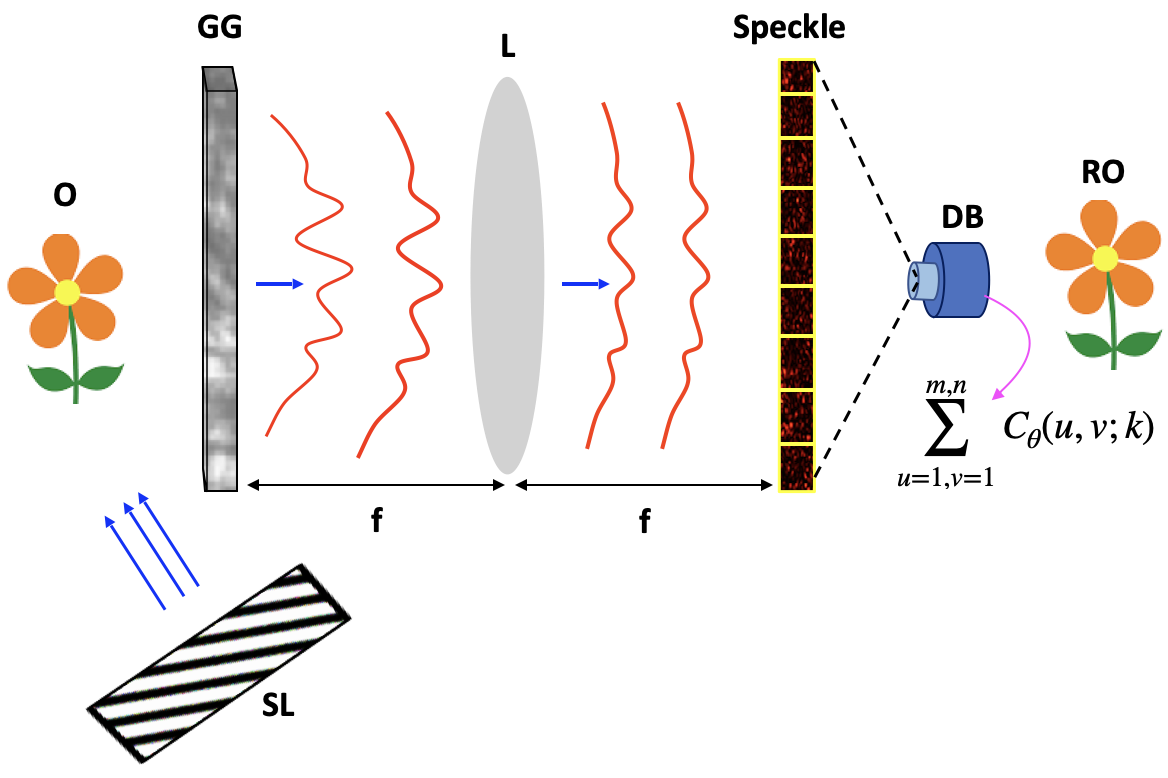} \\
    \small (b)
  \end{tabular}
  \caption{(a) Schematic representation of looking through scattering medium: PW- plane wave, O- object, GG- ground glass, L- a lens with focal distance $f$ (b) Schematic representation of looking through scattering medium with structured light: SL- structured light, RO- reconstructed object, DB- digital bucket detector.}
\end{figure}
The object immediately after the scattering layer is represented as
\begin{equation}\label{1}
        U(r)=p(r) \exp [i(\phi(r)+\varphi(r))]
\end{equation}
where $p(r)$ and $\varphi(r)$ are the amplitude and deterministic phase of the field at the scattering layer. The random phase introduced by the scatterer is represented as $\phi(r)$ and $r\equiv(x,y)$ denotes the position vector at the scattering plane. The random phase is considered to be uniformly distributed between phase values $[-\pi, \pi]$. The scattered field is Fourier transformed and as a result, the complex field at the observation plane becomes
\begin{equation}\label{2}
    E(\nu)=\int U(r) H(r, \nu) d r
\end{equation}
where $H(r, \nu)$ stands for the propagation kernel which is represented as $H(r, \nu)=\exp [-i \nu \cdot r]$ . The spatial frequency coordinate at the observation plane is represented by $\nu \equiv (u,v)$. The Fourier transform kernel used in Eq. \ref{2} supports in realization of the spatial ergodicity and stationarity at the observation plane.
The second-order correlation function of $E(\nu)$ of the speckle field at the observation plane is defined as\cite{takeda2014spatial}
\begin{equation}\label{3}
    \begin{array}{l}
F\left(\nu_{1}, \nu_{1}+\Delta \nu\right)=\left\langle E^{*}\left(\nu_{1}\right) E\left(\nu_{1}+\Delta \nu\right)\right\rangle \\
=\int E^{*}\left(\nu_{1}\right) E\left(\nu_{1}+\Delta \nu\right) d \nu_{1}=\int I(r) \exp [-i \Delta \nu \cdot r] d r
\end{array}
\end{equation}
Eq. \ref{3} represents the van Cittert-Zernike theorem obtained under the spatial averaging. A parenthesis $\langle . \rangle$ in Eq. \ref{3} represents the ensemble average which is replaced by the spatial averaging. This replacement is justified under consideration of spatially stationary and ergodic random field at the observation plane.
\\The fourth-order correlation function estimation is evaluated as
\begin{equation}\label{4}
    C\left(\nu_{1}, \nu_{1}+\Delta \nu\right)=\left\langle \Delta I\left(\nu_{1}\right) \Delta I\left(\nu_{1}+\Delta \nu\right)\right\rangle
\end{equation}
$I(\nu)= |E(\nu)|^{2}$ is intensity and $\Delta I(\nu)=I(\nu)-\langle I(\nu)\rangle$ is spatial fluctuation of the intensity over its mean value.\\
For a speckle pattern satisfying the Gaussian statistics, the fourth-order correlation can be represented in terms of the second-order correlation as
\begin{equation}\label{5}
    C(\Delta \nu) \propto|F(\Delta \nu)|^{2}
\end{equation}
Thus, appropriate optical geometry provides a squared modules of the complex coherence function. According to the van Cittert-Zernike theorem, as described in Eq. \ref{3}, the two-point complex correlation function of the random field can be used to reconstruct the object. On the other hand, right-hand side of Eq. \ref{5} is proportional to the far-field diffraction pattern, which is equivalent to replacing the amplitude of the field at the scattering layer by illumination function $I(r)= |p(r)|^{2}$. However, Eq. \ref{5} provides only amplitude of the complex coherence function, which is not sufficient for reconstruction of the object. The lost phase of $W(\Delta \nu)$  needs to be recovered and methods such as interferometry\cite{kumar2014recovery} or iterations\cite{das2017lensless} are useful for this purpose.\\

In this communication, we present an alternative approach for recovery of the complex coherence function and hence complex-valued object from the intensities of the speckle patterns. Idea is to use the structured light illumination at the object which is obstructed by a scattering layer as shown in Fig. 1(b). In the presence of structured light illumination, the source structure at the scattering layer is considered as $p(r)p_{\theta}(r,k)$, where $p_{\theta}(r, k)=a+b \cos (k \cdot r+\theta)$ represents the incident structured pattern. Term $a$ is un-modulated pattern, and $b$ represents the contrast. A two-dimension structured pattern with a spatial frequency $k \equiv (k_{x},k_{y})$  and transverse position coordinates $r \equiv(x,y)$  is projected over the scattering plane. As a result, for a given frequency $k$, the instantaneous field immediately after the scattering plane is represented as
\begin{equation}\label{6}
    U(r, k)=p(r) p_{\theta}(r, k) \exp [i(\phi(r)+\varphi(r))]
\end{equation}
The scattered field is Fourier transformed by a lens $L$ and an instantaneous complex field for a particular structured illumination pattern is represented as

\begin{equation}\label{7}
    E_{\theta}(\nu,k)=\int U(r,k) H(r, \nu) d r
\end{equation}
A two-dimensional (2D) intensity of size $M \times N$ pixels at the observation plane is captured and represented as 
\begin{equation}\label{8}
    I_{\theta}(u,v;k) = | E_{\theta}(u,v;k)|^{2}
\end{equation}
The 2D speckle pattern intensity is digitally stored in a personal computer (PC) and divided into discrete, non-overlapping, equal-sized small intensity patches $I_{\theta,s}(u,v;k)$ of size $m \times n$ pixels. Therefore we get total $S (= \frac {M\times N} {m\times n})$ number of incoherent intensity patches. Here, we plan to develop an alternative to the temporal averaging where random patterns of size $m \times n$ are created by a temporally fluctuating light. The random intensity variation from its mean value is expressed as
\begin{equation}\label{9}
\Delta I_{\theta,s}\left(u,v; k\right)=I_{\theta,s}\left(u,v;k\right)-\langle I_{\theta,s}\left(u,v;k\right)\rangle
\end{equation}
where a mean is represented as $\langle I_{\theta,s}\left(u,v;k\right)\rangle = \frac{\sum_{s=1}^{S}I_{\theta, s}(u, v ; k)}{S}$.
The cross-covariance of the intensities is given as
\begin{equation}\label{10}
    C_{\theta}(u,v;k)= \langle \Delta I_{\theta,s}\left(u,v; k\right)\Delta I_{\theta,s}\left(u,v; k\right)\rangle
\end{equation}
The 2D cross-covariance of the intensities is subjected to computational summation which corresponds to a spatially unresolved computational bucket detection and is represented as 
\begin{equation}\label{11}
    C_{\theta}(k)= \sum_{u=1, v=1}^{m,n}C_{\theta}(u,v;k)
\end{equation}
Eq. \ref{11} represents a spatially unresolved quantity. Learning from the idea of computational imaging with structured light illumination where stage of sampling the frequency spectrum is shifted from the camera to the programmable diffracting element. The complex Fourier spectrum is retrieved using following steps and with scaling of a variable $\nu \equiv k$.
A four-step phase shifting approach is introduced to recover a complex Fourier coefficient by combination of the $C_{\theta}(k)$ corresponding to  four illumination patterns, i.e. $p_{0}, p_{\pi/2}, p_{\pi},p_{3\pi/2}$. These four $C_{\theta}(k)$ $[\theta= 0, \pi/2, \pi, 3\pi/2]$ are combined to retrieve the Fourier spectrum as\cite{zhang2015single}
\begin{equation}\label{12}
\begin{array}{c}
    \left[C_{0}\left(k_{x}, k_{y}\right)-C_{\pi}\left(k_{x}, k_{y}\right)\right]+i\left[C_{\pi / 2}\left(k_{x}, k_{y}\right)-C_{3 \pi / 2}\left(k_{x}, k_{y}\right)\right]\\ \propto \int I(x, y) \exp \left[-i\left(k_{x} x+k_{y} y\right)\right] d x d y
    \end {array}
\end{equation}

Therefore, the Fourier coefficient $F(k_x,k_y)$ is recovered from the random intensity pattern as

\begin{equation}\label{13}
    F\left(k_{x}, k_{y}\right) \propto\left[C_{0}\left(k_{x}, k_{y}\right)-C_{\pi}\left(k_{x}, k_{y}\right)\right]+i\left[C_{\pi / 2}\left(k_{x}, k_{y}\right)-C_{3 \pi / 2}\left(k_{x}, k_{y}\right)\right]
\end{equation}
The desired complex field is recovered and object is reconstructed through scattering layer by computing Fourier coefficients $F\left(k_{x}, k_{y}\right)$ for a range of $(k_{x}, k_{y})$. 
\section{Experimental design and algorithm}

Fig. 2 depicts a possible experimental design of the proposed technique. A beam splitter (BS) folds a monochromatic collimated laser light which illuminates  a spatial light modulator (SLM). In Fig. 2, the SLM is a reflective type and it is loaded with $h$ number of sinusoidal gratings in a sequence.
\begin{figure}[!ht]
\centering\includegraphics[width=12cm]{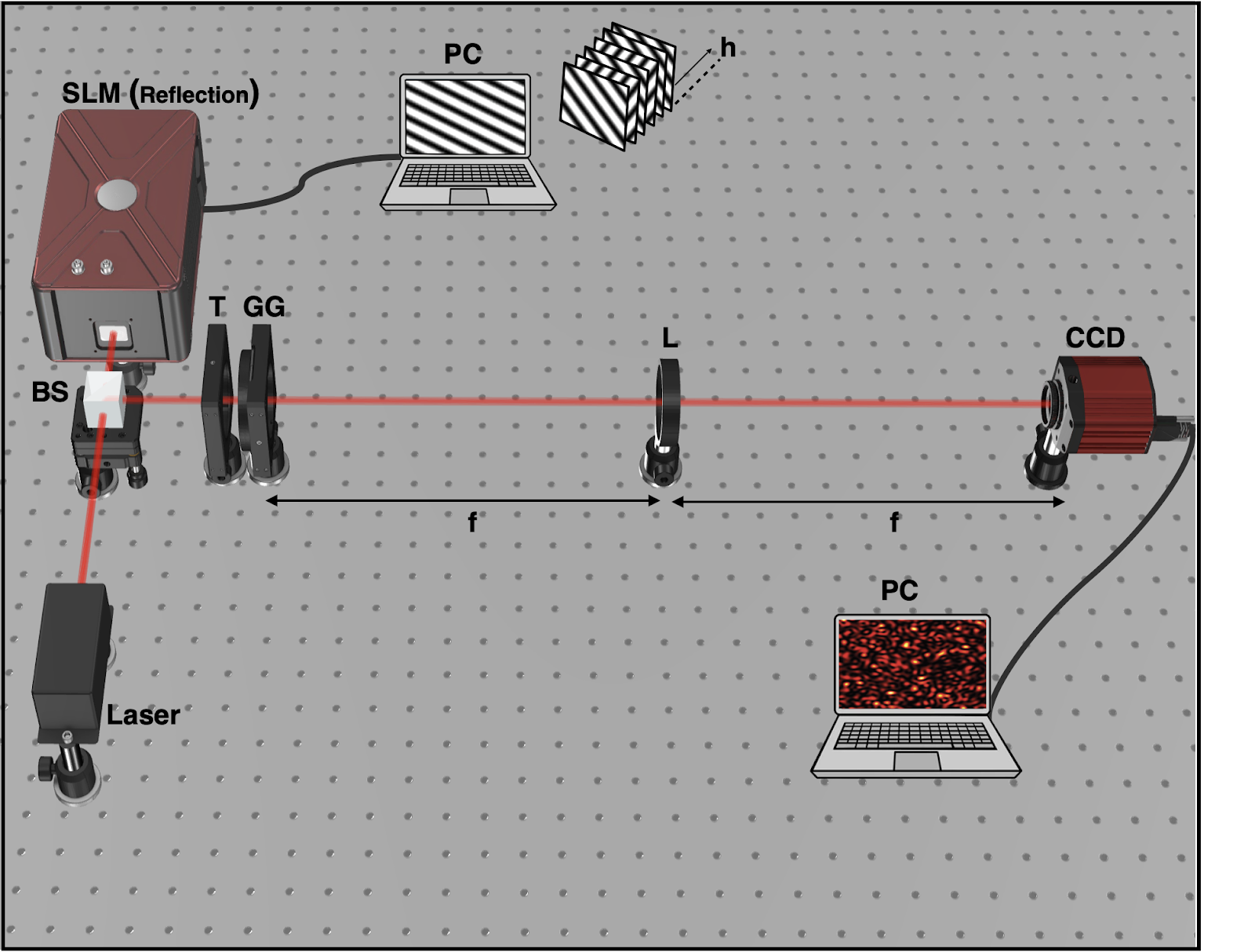}
\caption{Experimental configuration: T- transparency, BS- beam splitter, GG- ground glass, SLM- spatial light modulator, L- a lens with focal distance $f$, CCD- charge-coupled device detector, PC- personal computer.}
\end{figure}
This structured light is folded by the BS and illuminates a transparency $T$ which is placed in close vicinity of the scattering medium, i.e. a ground glass (GG). The GG generates a stochastic field with random phase variation and hence information of the transparency and structured light is scrambled into speckle pattern. The scattered light further propagates and is Fourier transformed at the charge-coupled device (CCD) plane by a lens $L$ . An random field at the CCD detector corresponds to the spatial frequency $(k)$ and at initial phase $\theta$ of the loaded structured pattern and represented by $ E_{\theta}\left(u, v ; k\right)$. The instantaneous two-dimensional (2D) speckle pattern intensity at the detector is represented as  $|E_{\theta}\left(u, v ; k\right)|^{2}$. We stored the intensity as a 2D matrix of dimension $M \times N$ in a personal computer (PC) for post-processing. These 2D pattern is divided into non-overlapping small patches $I_{\theta,s}(u,v;k)$ of size $(m \times n)$ and the cross-covariance of the intensities for a given frequency $(k)$ and initial phase $\theta$ is represented as $ C_{\theta}(k)$. We illuminate the transparency by the structured patterns with the full set of $(k)$ for a complete set of Fourier coefficients. A 4-step phase-shifting approach is used to extract each complex Fourier coefficient corresponding to $(k)$. In the Fourier domain, the number of Fourier coefficients is equal to the number of pixels in the reconstructed object.\\

The algorithm suggested in the literature is an iterative heuristic for reconstructing complex-valued object hidden behind the scattering medium from speckle patterns. The algorithm is designed on MATLAB and simulated on a PC. Fig. 3 shows the flow chart algorithm and steps are as following


\begin{enumerate}
    \item A  $M \times N$ pixels object is considered as transparency.
    \item Construction of sinusoidal patterns:  
    \begin{enumerate}
        \item Two-dimensional (2D) sinusoidal patterns of size $M \times N$ pixels with initial phase ${\theta}=(0,\pi/2,\pi,3\pi/2)$ are generated by using discretized spatial frequency space $-k\leq k_{x}, k_{y}\leq k$.
    \end{enumerate}
    \item Iterative steps:
    \begin{enumerate}
        \item For the particular frequency $(k)$, a 2-D sinusoidal pattern is generated and random light field is measured as expressed by Eq. \ref{8}
        
        \item The 2D speckle intensity of size $ M\times N$ is divided into discrete, non-overlapping, equal-sized patches of size $m \times n$ each. 
        
        \item Eq. \ref{9} is used to determine random intensity variations from its mean intensity. Eq. \ref{10} is used to compute the cross-covariance of the intensity for a spatial frequency $(k)$.
        
        \item Then the 2D cross-covariance of size $m \times n$ is integrated without a spatially resolved feature.
        
        \item Calculation of each complex valued Fourier coefficient $F\left(k_{x}, k_{y}\right)$ is extracted from four responses $C_{\theta}(\theta=0,\pi/2,\pi,3\pi/2)$ ( see Eq. \ref{13} ). 
    \end{enumerate}
We need to iterate the preceding processes throughout the discretized spatial frequency space $(k)$ until the last iteration is acquired to provide complete set of desired $ R\times R$ size Fourier coefficients. By considering the four-step phase-shifting process, a total of $4R^{2}(=4 \times R \times R)$ sinusoidal patterns were projected by the SLM. Therefore, a complex Fourier spectrum and object are recovered from the randomly scattered light.
    
\end{enumerate}
\begin{figure}[!ht]
\centering\includegraphics[width=12cm]{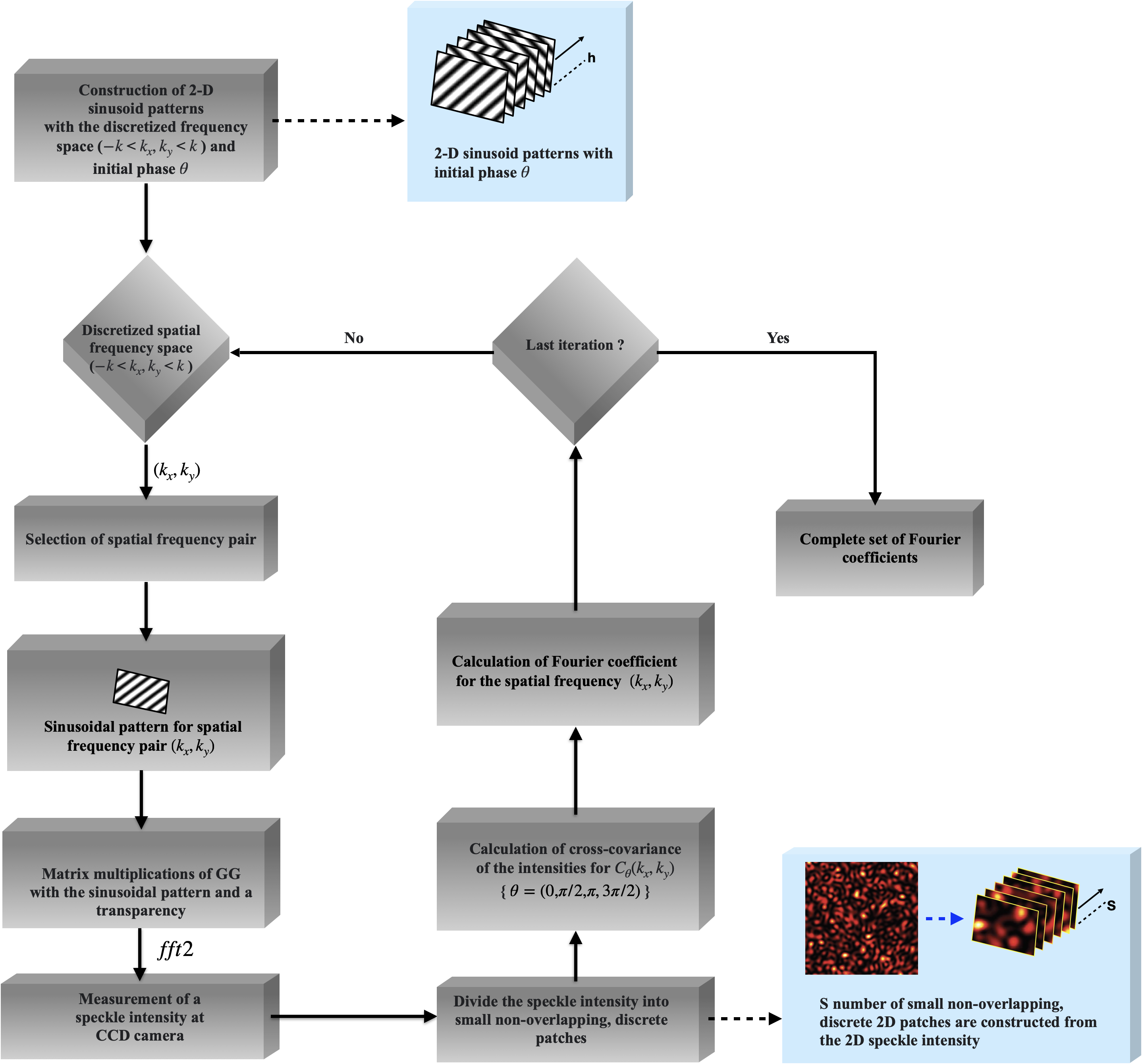}
\caption{Proposed algorithm steps.}
\end{figure}

\section{Results}
The cross-covariance and recovered complex-valued objects from the random light field are discussed and presented in this section. To demonstrate utility of our technique in looking through diffuser, we considered two different types of transmittance, namely off-axis and in-line holograms. 
\subsection{Reconstruction of off-axis hologram through diffuser :}
A computer-generated off-axis hologram is used as a transparency. An off-axis object is Fourier transformed and coherently interfered with a reference to generate an off-axis computer-generated hologram (CGH). This hologram is read out with structured light and propagates through GG. Using the structured light and proposed correlation approach, reconstructed complex fields result of two different objects \enquote{8} and \enquote{Trishula} are shown in Figs. 4 and 5.
 In each figs. 4(i)(a-d), 5(i)(a-d) four measured speckle patterns of size $1000 \times 1000$ for a particular spatial frequency $(k)$ and $\theta =(0,\pi/2,\pi,3\pi/2)$ are shown. For the spatial averaging perpose, these speckle patterns are devided into $100$ small non-overlapping, discrete, equal-sized patches of size $100 \times 100$ and are shown as small yellow bordered squares. Figs. 4(ii) and 5(ii) are recovered off-axis Fourier hologram using the proposed method.\\
A two-dimensional Fourier transform of the recovered transparency show three spectra emerge:  a non-modulating central DC term, a desired off-axis spectrum, and it's conjugate.  Off-axis location of the reconstructed complex-valued objects depends on the carrier frequency used in the CGH. 
Structured illumination patterns $p_{\theta}(r,k)$ are generated by using a=0.5, b=0.5, and the spatial frequencies range is $-1.5\leq(k_{x},k_{y})\leq1.5$ at increments of 0.03, in order to reconstruct  complex valued object of size $100 \times 100$ through scattering medium. Reconstructed amplitude and phase of both objects are illustrated in Figs. 4(iii),(iv) and 5(iii),(iv) and the undesirable DC terms with high-frequency content are digitally suppressed to highlight the objects.

\subsection{Reconstruction of in-line hologram through diffuser :}
In another demonstration, we test our method to retrieve complex-valued object by reconstructing in-line CGH through diffuser.
An inline hologram is used as transparency which is digitally recorded by considering a Gabor's geometry of recording.
This in-line hologram is read out with structured light through diffuser. Digital reconstruction of an inline hologram suffers from issue of twin image problem and this can be removed by using iterations\cite{kim2019imaging} and deep learning(DL)\cite{li2020deep}. Using algorithm steps of our proposed technique as explored earlier, the in-line hologram is recovered from the random speckle pattern as shown in Fig. 6. The reconstruction results of object \enquote{A} were shown in Fig. 6. In figs. 6(i)(a-d) four
measured speckle patterns of size $1000 \times 1000$ for a particular spatial frequency $(k)$ and $\theta =(0,\pi/2,\pi,3\pi/2)$ are shown. For the spatial averaging purpose, these speckle patterns are divided into $100$ small non-overlapping, discrete, equal-sized patches of size $100 \times 100$ and are shown as small yellow bordered squares. In order to reconstruct complex-valued object of size  $400 \times 400$ through scattering medium, structured illumination patterns $p_{\theta}(r,k)$ are formed with a=0.5, b=0.5, and the spatial frequencies range is $-2.5\leq(k_{x},k_{y})\leq2.5$ at increments of 0.0125. 6(ii) is recovered in-line hologram.\\ 
The reconstructed in-line hologram is backpropagated to the object plane of distance of 300 mm. Reconstructed phase and amplitude of object \enquote{A} is shown in Figs. 6(iii),(iv) respectively. It can be seen that the reconstruction suffers from the superimposed out-of-focus twin-images. Then the out-of-focus twin-image formed in the recovered digital in-line hologram is then eliminated using a deep network with an encoder-decoder (also called auto-encoder)\cite{li2020deep} type architecture. This method employed a blind single-shot hologram reconstruction based just on the captured sample, eliminating the requirement for a large dataset of samples with available ground truth to train the model, which takes the advantage of the main characteristic of auto-encoders. As their method used single-shot imaging which eliminated the need for sophisticated experimental setup of multi-height phase retrieval and time-consuming DL-based methods, which makes it cost-effective and computationally feasible. Figs. 7 (i), (ii) show the phase and amplitude distribution of object \enquote{A} where twin-image artifacts are completely removed.

\begin{figure}[!ht]
\centering\includegraphics[width=12cm]{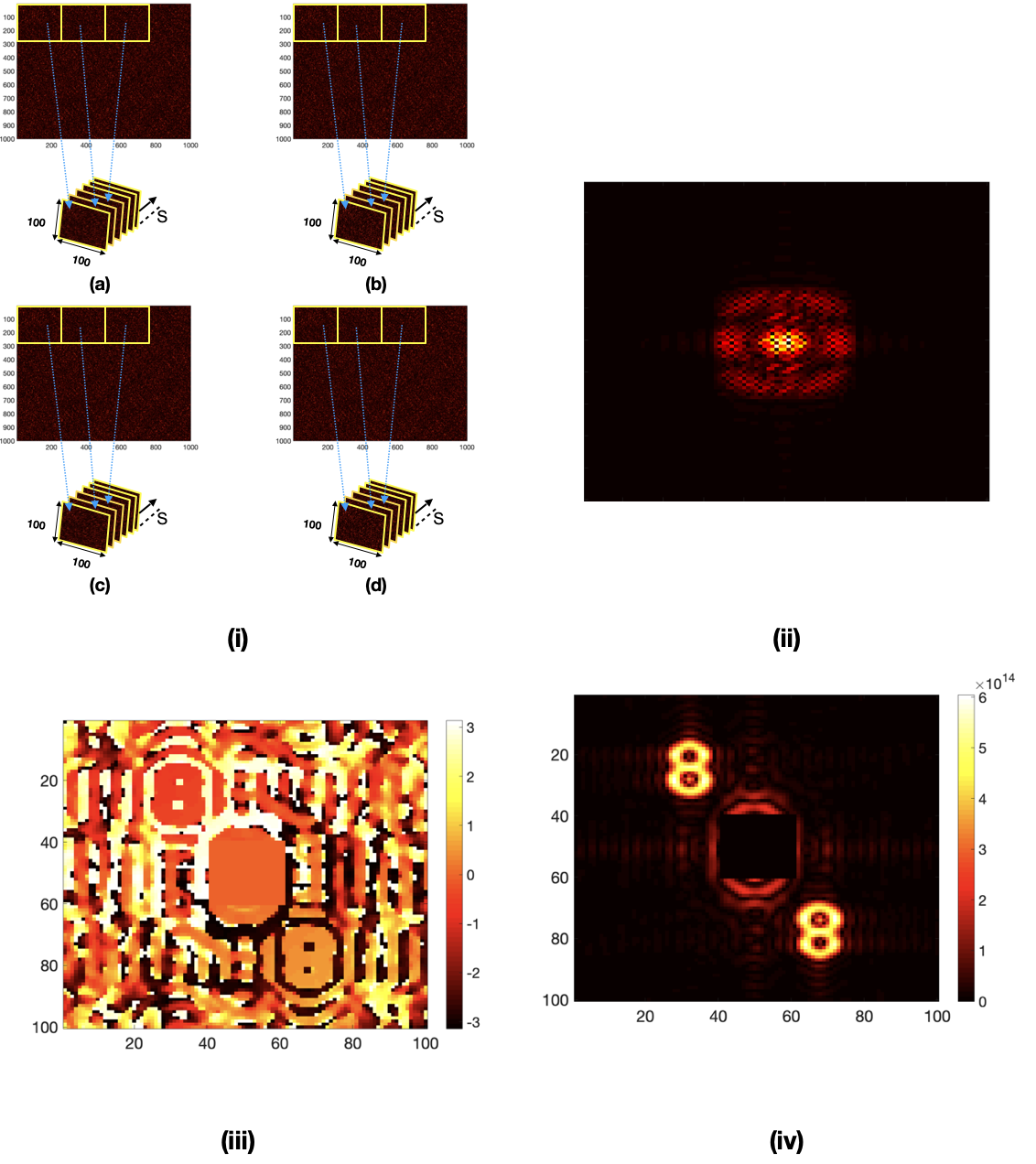}
\caption{Simulated results for object \enquote{8} (i) measured speckle patterns [a-d] corresponding to four phases $[0,\pi/2,\pi,3\pi/2]$ (ii) recovered off-axis hologram (iii) \& (iv) recovered phase and amplitude distribution.}
\end{figure}
\begin{figure}[H]
\centering\includegraphics[width=12cm]{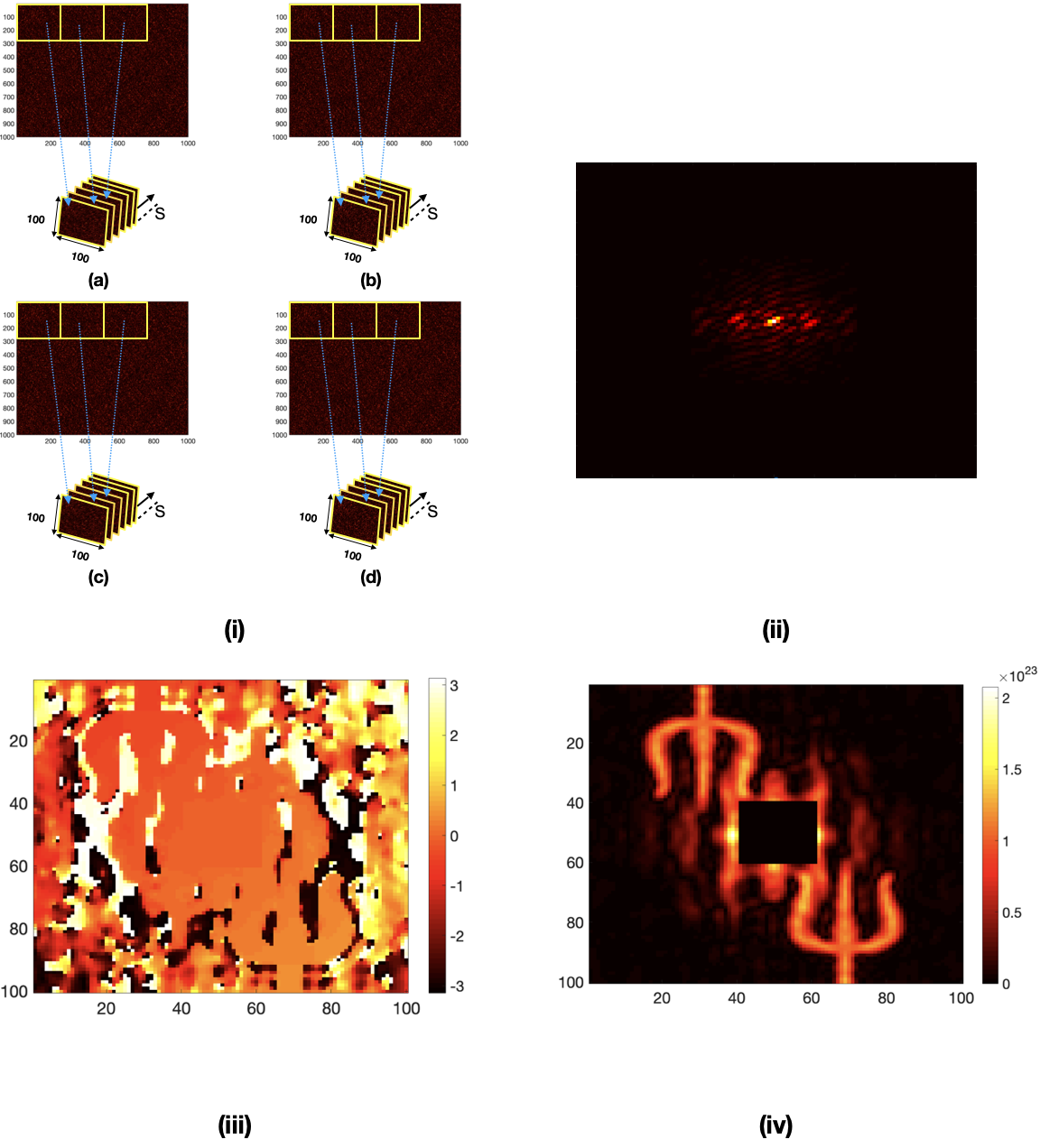}
\caption{Simulated results for object \enquote{Trishula} (i) measured speckle patterns [a-d] corresponding to four phases $[0,\pi/2,\pi,3\pi/2]$ (ii) recovered off-axis hologram (iii) \& (iv) recovered phase and amplitude distribution.}
\end{figure}

\begin{figure}[H]
\centering\includegraphics[width=12cm]{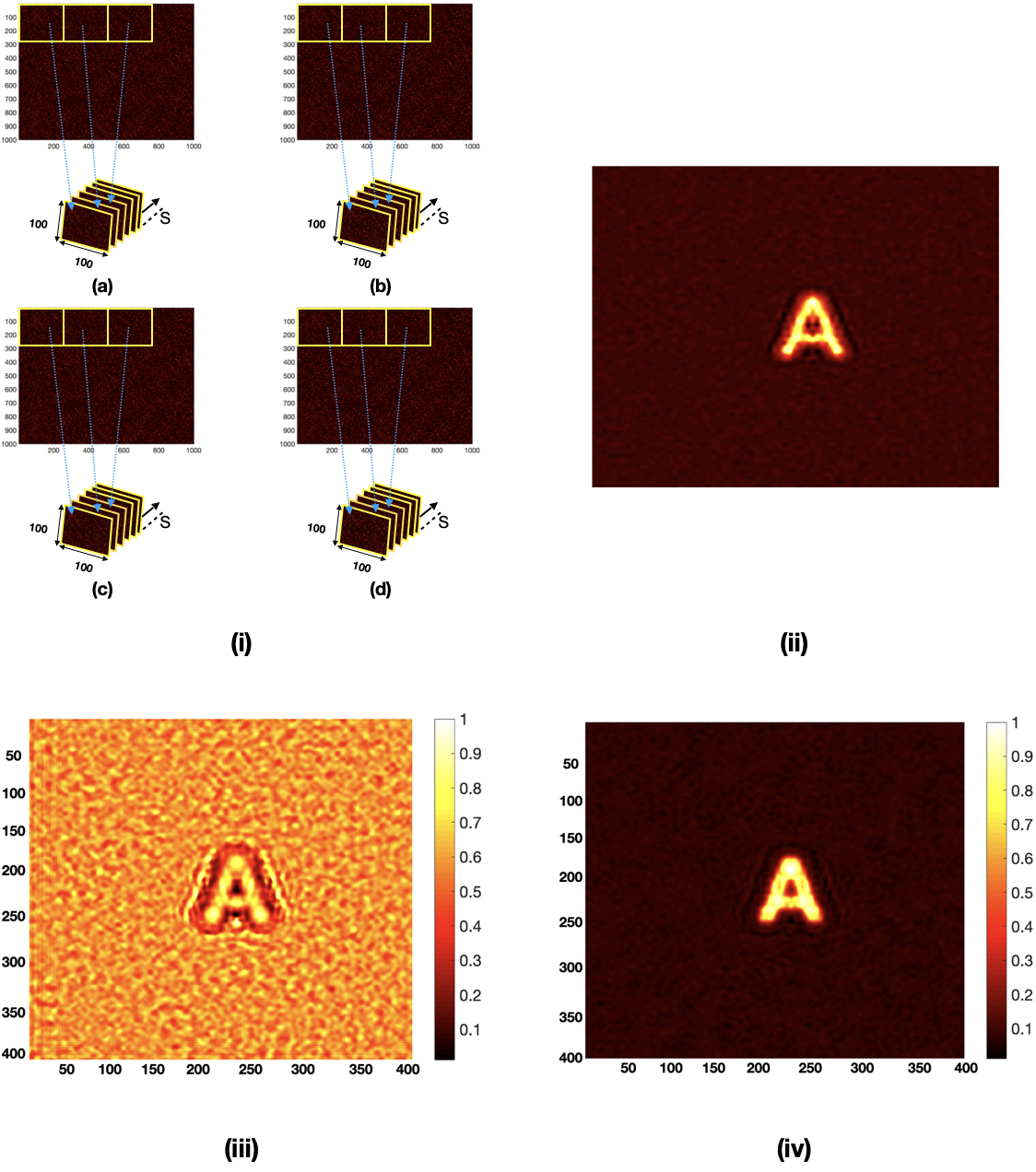}
\caption{Simulated results for object \enquote{A} (i) measured speckle patterns [a-d] corresponding to four phases $[0,\pi/2,\pi,3\pi/2]$ (ii) recovered in-line hologram (iii) \& (iv) back propagated phase and amplitude distribution respectively in presence of twin-image}
\end{figure}
\begin{figure}[H]
\centering\includegraphics[width=12cm]{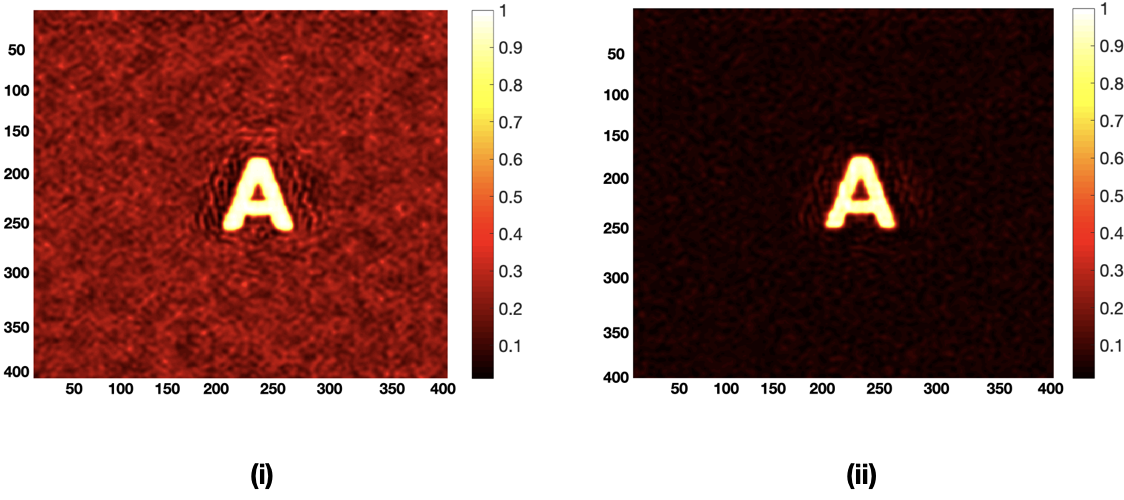}
\caption{ (i) \& (ii) Reconstructed phase and amplitude distributions respectively with twin-image removed by auto-encoder model}
\end{figure}

\section {Conclusion}
In summary, we have proposed a new method for reconstruction of complex field through random scattering medium. This technique utilizes the structured light illumination and is capable to recover the complex field through the intensity correlations in a non-interferometric approach. Furthermore, proposal to devise a new scheme with spatially fluctuating field to implement intensity correlation with computational bucket detector provides a new way to apply the correlation-based imaging techniques.The technique is expected to find applications in imaging with random light, quantitative phase imaging etc.
\section{Declaration of Competing Interest}
The authors declare that they have no known competing financial interests or personal relationships that could have appeared to influence the work reported in this paper.
\section{CRediT authorship contribution statement}
\textbf{Aditya Chandra Mandal}: Conceived the idea, Investigation, Simulation, Data analysis, Writing – original draft.
\textbf{Manisha}: Ideas, Data analysis, Writing – original draft.
\textbf{Abhijeet}: Simulation and data analysis.
\textbf{Zeev Zalevsky}: Methodology, Revision.
\textbf{Rakesh Kumar Singh}: Ideas, Formulation of research goals, Revision and editing, Funding acquisition, Supervision.
\section{Acknowledgments}
The work is supported by the Science and Engineering Research Board (SERB) India- CORE/2019/000026 and Department of Biotechnology (DBT)- BT/PR35557/MED/32/707/2019.
\section*{References}

\bibliographystyle{elsarticle-num}
\bibliography{elsarticle-template}
\end{document}